\documentclass[aps,prb,superscriptaddress,twocolumn,showpacs,preprintnumbers,amsmath,amssymb]{revtex4}

\usepackage{array}
\usepackage{amsmath}
\usepackage{amssymb}
\usepackage{epsfig}
\usepackage{graphicx}
\usepackage{bbm}

\newcommand{\Ref}[1]{Ref.~\onlinecite{#1}}
\newcommand{\bs}[1]{{\boldsymbol#1}}

\newcommand{\ie}{{\emph{i.e.~}}}
\makeatletter

\newcommand{\Rmnum}[1]{\expandafter\@slowromancap\romannumeral #1@}
\makeatother
\newcommand{\imth}{\hspace{1pt}\mathrm{i}\hspace{1pt}}

\newcommand{\eg}{{\emph{e.g.~}}}

\newcommand{\plash}[1]{#1\!\!\!/}

\begin{document}
\title{$Z_2$ spin liquid and chiral antiferromagnetic phase in Hubbard model on the honeycomb lattice:
duality between Schwinger-fermion and Schwinger-boson
representations}

\author{Yuan-Ming Lu}\author{Ying Ran}
\affiliation{Department of Physics, Boston College, Chestnut Hill,
MA 02467, USA}
\date{\today}

\begin{abstract}
In our previous work\cite{Lu2010}, we identify the
Sublattice-Pairing State (SPS) in Schwinger-fermion representation
as the spin liquid phase discovered in recent numerical study on a
honeycomb lattice\cite{Meng2010}. In this paper, we show that SPS is
identical to the zero-flux $Z_2$ spin liquid in Schwinger-boson
representation found by Wang\cite{Wang2010} by an explicit duality
transformation. SPS is connected to an \emph{unusual}
antiferromagnetic ordered phase, which we term as
chiral-antiferromagnetic (CAF) phase, by an $O(4)$ critical point.
CAF phase breaks the $SU(2)$ spin rotation symmetry completely and
has three Goldstone modes. Our results indicate that there is likely
a hidden phase transition between CAF phase and simple AF phase at
large $U/t$. We propose numerical measurements to reveal the CAF
phase and the hidden phase transition.
\end{abstract}


\maketitle

\textit{\textbf{Introduction} } In a recent numerical
study\cite{Meng2010}, it is found that the Hubbard model on the
honeycomb lattice hosts a spin disordered insulating phase in the
neighborhood of the Mott transition, which does not break any
physical symmetry. If this is the case, this phase should be a novel
spin liquid phase with fractionalized
excitations\cite{Oshikawa2006}. Similar fractionalized spin liquids
have been shown to exist in various artificial
models\cite{Affleck1988a,Read1991,Rokhsar1988,Moessner2001,Wen2003,Kitaev2006},
but so far there is no simple and hopefully realizable Hamiltonian
that hosts such exotic phases. This remarkable numerical study makes
an important footstep along finding spin liquid phases in correlated
electron systems.

There are a lot of different spin liquids on the honeycomb lattice, characterized by different topological orders, or different Projective Symmetry Groups (PSG)\cite{Wen2002}.
For example, in a previous study we show that there can be 128
distinct spin liquid phases within the Schwinger-fermion
representation\cite{Lu2010}. Which one is realized in the Hubbard
model? In the numerical study\cite{Meng2010}, it is shown that the
spin liquid phase have a full energy gap, and is smoothly connected
(\ie through a continuous phase transition) to both the semi-metal
phase for small $U/t$ and the Neel phase for large $U/t$. These
three conditions strongly restricts the candidate spin liquid
phases.

In our previous study by Schwinger-fermion approach\cite{Lu2010}, we
use only two of the three conditions and show that there is only one
natural spin liquid, coined the Sublattice Pairing State(SPS), which
has a full energy gap and can be smoothly connected to the
semi-metal phase. Is SPS compatible with the third condition? In
other words, can SPS be connected to a magnetic ordered phase by a
continuous phase transition? Describing ordinary magnetic ordered
phase in Schwinger-fermion approach has been a puzzle for a long
time. In this work we show a solution, which allows us to study
phase transitions between SPS and a magnetic ordered phase.

\begin{figure}
 \includegraphics[width=0.23\textwidth]{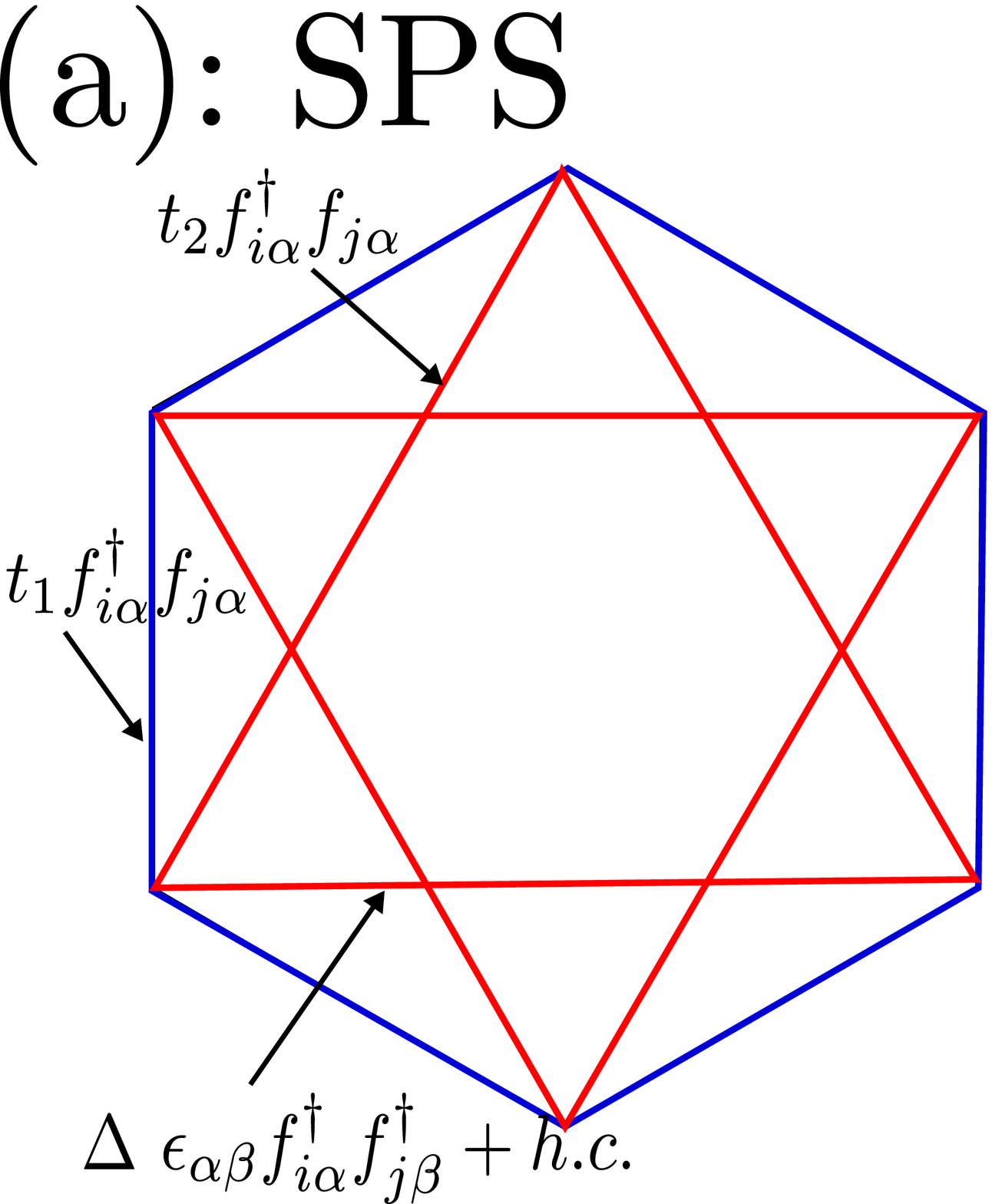}\;\;\;\;\;\includegraphics[width=0.23\textwidth]{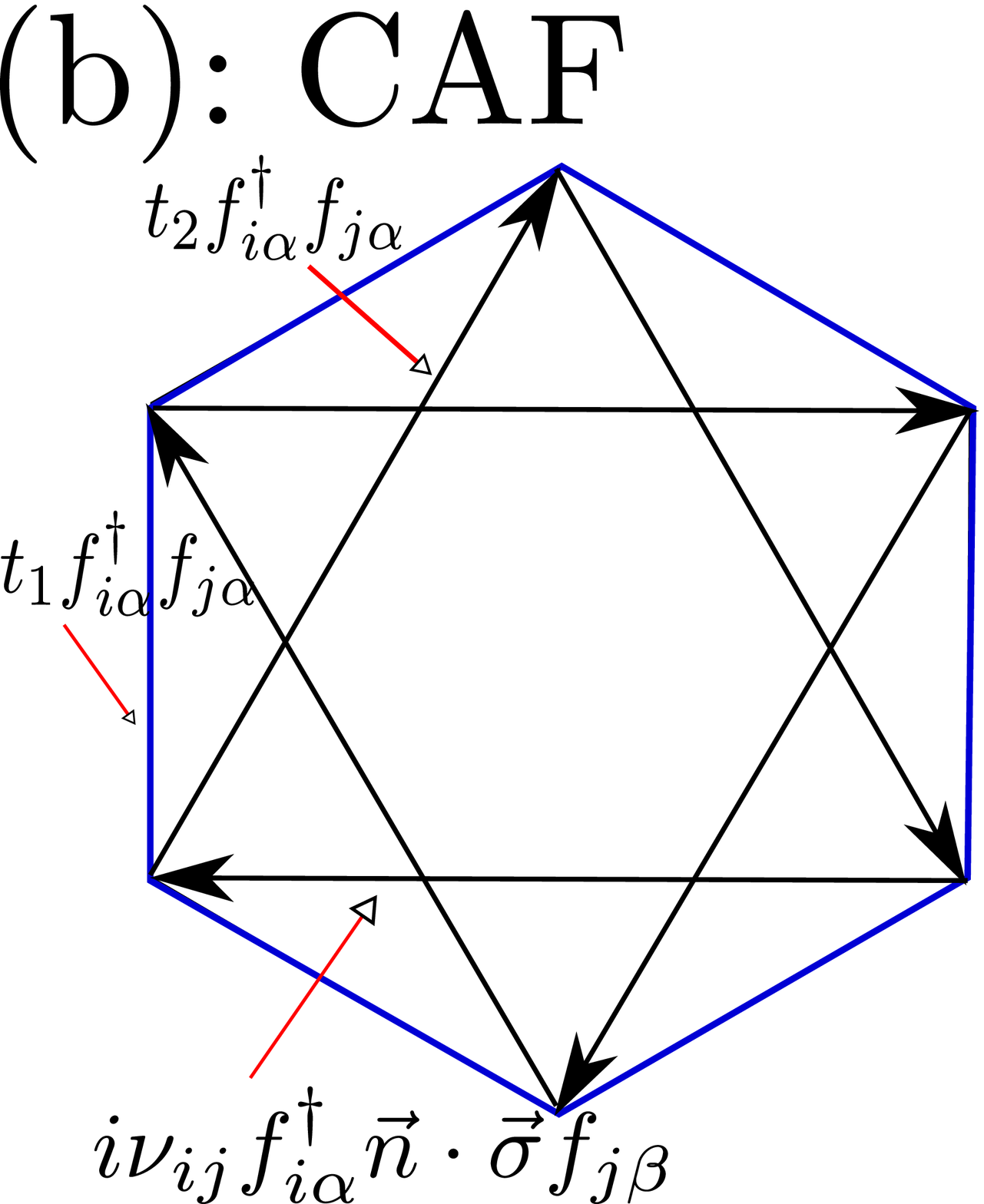}
\caption{(color online) Mean-field ansatz of (a) SPS phase and (b)
CAF phase in terms of $f$-fermion. $t_{1,2}$ are real
fermion-hopping parameters, while s-wave pairing order parameter
$\Delta$ can be complex. Vector $\vec{n}$ is the QSH order parameter
in the CAF phase. $\nu_{ij}=1$ if $i\rightarrow j$ is along the
arrow direction.} \label{fig:sps+caf}
\end{figure}

SPS is a fully-gapped $Z_2$ spin liquid on the honeycomb lattice.
Its mean-field fermionic spinon band structure, after choosing a
proper gauge, is given as follows: (see FIG. \ref{fig:sps+caf}a)
\begin{align}
 H^{MF}_{SPS}&=t_1\sum_{<ij>}f_{i\alpha}^{\dagger}f_{j\alpha}+t_2\sum_{<<ij>>}f_{i\alpha}^{\dagger}f_{j\alpha}\notag\\
&-\mu\sum_if_{i\alpha}^{\dagger}f_{i\alpha}+\Delta\sum_{<<ij>>}\epsilon_{\alpha\beta}f_{i\alpha}^{\dagger}f_{j\beta}^{\dagger}+h.c.\label{eq:SPS_mf}
\end{align}
where $t_{1,2}$ are real numbers. In Schwinger-fermion approach,
$f$-spinons are coupled to an $SU(2)$ gauge
field\cite{Affleck1988,Wen2002}. However due to non-zero $t_2$ and
$\Delta$, this $SU(2)$ gauge degree of freedom is reduced to $Z_2$
through Higgs mechanism. Thus at low energy $f$-spinons in
Eq.(\ref{eq:SPS_mf}) are coupled to a dynamical $Z_2$ gauge field
and stay in the deconfined phase.

Recently Wang\cite{Wang2010} identified the 0-flux state as the most
promising spin liquid phase in the Schwinger-boson approach. 0-flux
state is also a $Z_2$ state with a full energy gap, smoothly
connected to an antiferromagnetic phase through an $O(4)$ critical
point. However it is not clear whether the 0-flux state can be
smoothly connected to the semi-metal phase. Can be SPS related to
the 0-flux state? The two states are described in two completely
different languages: one in Schwinger-fermion and the other in
Schwinger-boson, and the relation between these two representations
has been a longstanding problem. In this paper, we find that
strikingly, SPS and 0-flux state are identical by an explicit
duality transformation in the low energy effective theory.

\begin{figure}
 \includegraphics[width=0.4\textwidth]{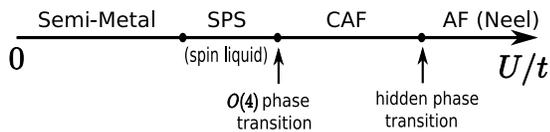}
\caption{Proposed schematic phase diagram of the Hubbard model on
the honeycomb lattice. The hidden phase transition might happen at a
large $U/t$, which is not reached in the numerical
study\cite{Meng2010}.} \label{fig:phase_diagram}
\end{figure}

We also find out that the antiferromagnetic phase connected to the
0-flux state (or SPS) is rather unusual and \emph{not} the simple
Neel phase, because it breaks the $SU(2)$ spin-rotation symmetry
completely and has three Goldstone modes. On the other hand the
magnetic order is still colinear. We dub this phase as
chiral-antiferromagnetic(CAF) phase. In CAF phase, aside from the
usual antiferromagnetic spin order parameter $\vec N=(-1)^{i_s} \vec
S_i$ where $i_s=0,1$ for A and B sublattices respectively, there is
another vector-chirality spin order parameter $\vec
n=\sum_{<<ij>>}\nu_{ij}\vec S_{i}\times\vec S_{j}$ whose expectation
value satisfies $\langle\vec n\rangle\perp\langle\vec N\rangle$, and
$\nu_{ij}=+1$$(-1)$ if one makes a left(right) turn when going from
site $j$ to $i$ as shown by the arrows in FIG. \ref{fig:sps+caf}b.
Since the usual AF phase should exist in the large $U/t$
limit\cite{Gros1987}, our results suggest a hidden phase transition,
which might happen in the "Neel" ordered phase of the numerical
study\cite{Meng2010} or at larger $U/t$ not studied before. In other
words the numerical study may not distinguish CAF and simple AF
phases. We propose the schematic phase diagram as shown in
Fig.\ref{fig:phase_diagram}.

\textit{\textbf{Continuous phase transition from SPS to CAF phase}}
We first discuss the continuous phase transition from SPS to CAF
phase. Describing an ordinary magnetic ordered phase in
Schwinger-fermion approach is highly non-trivial: this is because a
Schwinger-fermion mean-field ansatz has at least an unbroken $Z_2$
gauge symmetry and to describe a regular magnetic phase, the gauge
degree of freedoms must be confined. The first demonstration of a
regular magnetic-ordered phase in Schwinger-fermion approach is
given in \Ref{Ran2008}. It is shown that the easy-plane
antiferromagnetic order (XY order) on the honeycomb lattice is
described by a quantum spin Hall (QSH) band structure of the
fermionic spinon along $S^z$ direction coupled with a dynamical
$U(1)$ gauge field, where $\uparrow (S^z=\frac12)$ spinon band has a
Chern-number $\nu=1$ while $\downarrow (S^z=-\frac12)$ spinon band
has a Chern-number $\nu=-1$. Because of the QSH effect, the gauge
fluctuation is bound to $S^z$ spin density fluctuation, and the
Goldstone mode of the easy-plane Neel order is nothing but the
photon of the $U(1)$ gauge field. The long-range spin-spin
correlation function in the Neel phase is dual to the long-range
monopole-monopole correlation function in the Coulomb phase of the
$U(1)$ gauge field. Indeed one can compute the monopole quantum
number\cite{Ran2008} to show the spin order pattern is
antiferromagnetic.

\begin{figure}
 \includegraphics[width=0.3\textwidth]{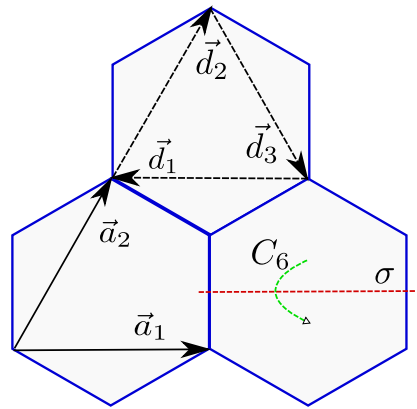}
\caption{The honeycomb lattice and its Bravais lattice vector $\vec
a_{1,2}$. $\vec d_{1,2,3}$ are the three vectors defined in the
effective Lagrangian of $z$ boson for the SPS-CAF phase transition.
Two generators of symmetry group are also shown: $60^\circ$ rotation
$C_6$ around the center of a plaquette and the horizontal mirror
reflection $\sigma$.} \label{fig:honeycomb}
\end{figure}

Armed with this result, we now 
consider an $SU(2)$ spin rotation symmetric system. CAF phase is
described by spinon band struture in the presence of a fluctuating
$O(3)$ QSH order parameter $\vec n$, coupled with a $U(1)$ gauge
field. Its mean-field ansatz is: (see FIG. \ref{fig:sps+caf}b)
\begin{align}
H^{MF}_{CAF}&=t_1\sum_{<ij>}f_{i\alpha}^{\dagger}f_{j\alpha}+t_2\sum_{<<ij>>}f_{i\alpha}^{\dagger}f_{j\alpha}\notag\\
&-\mu\sum_if_{i\alpha}^{\dagger}f_{i\alpha}+\vec
n\cdot\sum_{<<ij>>}i\nu_{ij}f_{i\alpha}^{\dagger}\vec
\sigma_{\alpha\beta}f_{j\beta},\label{eq:CAF_mf}
\end{align}
Note that the non-vanishing $t_2$ term, as well as the QSH order $\vec n$, breaks the $SU(2)$ gauge
symmetry down to $U(1)$.

In the Dirac limit ($t_2,|\vec n|\equiv m\ll t_1$) we can write down
the effective Lagrangian of CAF phase in imaginary time:
\begin{align}
\mathcal{L}=&\psi^{\dagger}\gamma_0(\partial_{\mu}-ia_{\mu})\gamma_{\mu}\psi+m\;\hat n \cdot \psi^{\dagger}\vec M\psi\notag\\
&+\frac{1}{4g_a^2}f_{\mu\nu}^2+\frac{1}{2u}(\partial_{\mu}\hat n)^2.
\label{eq:CAF_eff}
\end{align}
where $\psi$ is a 8-component complex fermion describing the long
wavelength part of $f$-spinon around the two Dirac points $K$ and
$K'=-K$ in the Brillouin Zone:
$\psi_k\equiv(f_{K+k,A},f_{K+k,B},f_{K'+k,B},f_{K'+k,A})$ and we
neglect spin indices. $\gamma_{\mu}$ are Pauli matrices in the
sublattice space: $\gamma_0=\mu_3,\gamma_1=\mu_2,\gamma_2=-\mu_1$.
$\vec M=\gamma_0\vec\sigma$ are the QSH masses, and space/time are
rescaled so that the Fermi velocity is one. Non-universal coupling
$u$ describes the fluctuation of the QSH order parameter: unit
vector $\hat n$.

There are three gapless modes in CAF phase: two $\hat n$ fluctuating
modes and one photon mode. The photon mode is in-plane spin wave of
anti-ferromagnetic order $\vec N$ ($\vec N\perp
\hat n$), and the spin $SU(2)$
symmetry is completely broken. The combination $C_6\cdot\bs{T}$ of
the $60^{\circ}$ space rotation around the hexagon center and
time-reversal $\bs T$ leaves both order parameters invariant. This
symmetry indicates that the magnetic order in CAF phase is still
collinear.

Comparing Eq.(\ref{eq:CAF_mf}) with Eq.(\ref{eq:SPS_mf}), s-wave
pairing $\Delta$ of spinons in SPS phase is replaced by the $O(3)$
QSH order $\vec n$ in CAF phase. If we group these orders together
into a 5-component vector $\vec V=(\mbox{Re}\Delta, \mbox{Im}\Delta,
\vec n)$ and ignore gauge fields for the moment, as pointed out by
Grover and Senthil\cite{Grover2008}, fluctuations of $\vec V$ has a
Wess-Zumino-Witten (WZW) term after integrating out fermions. This
WZW term allows us to construct an $O(4)$ phase transition between
SPS and CAF.

WZW term is a topological Berry phase in the non-linear sigma model
of $\hat V=(v_1,v_2,v_3,v_4,v_5)$ ($|\hat V|=1$) in 2+1
dimension\cite{Abanov2000}:
\begin{align}
 S_{WZW}=\frac{2\pi\imth}{Area(S^4)}\int_0^1 d\rho\int d^3 x\epsilon_{abcde}v_a\partial_{\rho}v_b\partial_{t}v_c\partial_{x}v_d\partial_{y}v_e\label{eq:WZW}
\end{align}
where $Area(S^4)=\frac{8\pi^2}{3}$, and $\hat V(x,y,t,\rho)$ is a
smooth extension of space-time configuration $\hat V(x,y,t)$ to
4-disk: $\hat V(x,y,t,\rho=1)=\hat V(x,y,t)$ and $\hat
V(x,y,t,\rho=0)=\hat V_0$ is a fixed vector. The physical meaning of
the WZW term is that a Skyrmion (anti-Skyrmion) of $\hat n$ in two
space dimensions actually carries fermion charge $2(-2)$. And the
hedgehog instanton of $\hat n$ in 2+1 dimension creates a charge-2
s-wave fermion pair. Due to this WZW term, as shown in
\Ref{Grover2008}, a direct phase transition between a QSH insulator
and an s-wave superconductor on the honeycomb lattice becomes
possible.

Let us keep the WZW term in mind and study the CAF to SPS phase
transition. Starting from the effective theory of the CAF phase
Eq.(\ref{eq:CAF_eff}), it is convenient to introduce the $CP^1$
representation of the $\hat n$ order parameter: $\hat n=
w^{\dagger}\vec \sigma w$ where $w=(w_1,w_2)^T$ are two complex
numbers satisfying $|w_1|^2+|w_2|^2=1$. This representation has a
$U(1)$ gauge redundancy and thus $w$-bosons couple to a $U(1)$ gauge
field $A_{\mu}$ at low energy. After integrating out the $f$-spinon
of Eq.(\ref{eq:CAF_eff}) the effective Lagrangian is:
\begin{align}
L=&|(\partial_{\mu}-\imth A_{\mu})w|^2+r|w|^2+s |w|^4+\frac{1}{2g_a^2}f_{\mu\nu}^2\notag\\
&+\frac{1}{2g_A^2}F_{\mu\nu}^2+\frac{\imth}{\pi}\epsilon_{\mu\nu\lambda}A_\mu\partial_\nu
a_\lambda\label{eq:eff_lag}
\end{align}
where the constraint $|w_1|^2+|w_2|^2=1$ is softened\footnote{The
constraint can be enforced by a Lagrangian multiplier $\lambda$.
Eq.(\ref{eq:eff_lag}) can be obtained by introducing the saddle
point expansion of $\lambda$.}. $f_{\mu\nu}=\partial_\mu
a_\nu-\partial_\nu a_\mu$ and $F_{\mu\nu}=\partial_\mu
A_\nu-\partial_\nu A_\mu$ are the field strengths of the two gauge
fields. A key observation of the present work is the mutual
Chern-Simons (CS) term between $A_{\mu}$ and $a_{\mu}$. While an
explicit derivation of this term is given in Appendix
\ref{app:mutual_CS}, it can be easily understood: a Skyrmion of
$\hat n$ is well-known to be represented as a $2\pi$ $A_{\mu}$ gauge
flux. Now the WZW term dictates this $2\pi$ $A_{\mu}$ flux to carry
charge-2 of $a_{\mu}$ gauge field, which is exactly described by the
mutual CS term.

What are the phases described by the effective Lagrangian
Eq.(\ref{eq:eff_lag})? When $r<0$, $w$-boson condenses and $\hat n$
is ordered, so the system is in the CAF phase. Note that in this
phase the mutual Chern-Simons term does not qualitatively modify the
low energy gauge dynamics: the $A_{\mu}$ gauge fields are gapped out
due to the Higgs mechanism. 
Remarkably, when $r>0$, $w$-bosons are gapped and $\hat n$ is
disordered, and the system is actually in the $Z_2$ SPS spin liquid
phase. The $U(1)$ mutual Chern-Simons term or the WZW term is
essential to make this identification. The relation between mutual
CS theory and $Z_2$ gauge theory was firstly discussed in
\Ref{Kou2008}. Here, based on the WZW term and monopole physics, we
are able to further identify the PSG of this $Z_2$ state.

In the $\hat n$ disordered phase, the mutual CS term opens
up mass gaps for both $A_{\mu}$ and $a_{\mu}$ gauge
fluctuations\cite{Kou2008}: $m_a^2\sim g_A^2,m_A^2\sim g_a^2$.
Meanwhile, the real time (Minkowski spacetime) equations of motion
(with source term $-a^\mu j_\mu-A^\mu J_\mu$) become
\begin{align}
 j_{\mu}&=\frac{1}{g_a^2}\partial^{\nu}f_{\nu\mu}+\frac{1}{\pi}\epsilon_{\alpha\beta\mu}\partial^{\alpha}A^{\beta}\notag\\
J_{\mu}&=\frac{1}{g_A^2}\partial^{\nu}F_{\nu\mu}+\frac{1}{\pi}\epsilon_{\alpha\beta\mu}\partial^{\alpha}a^{\beta}+iw^{\dagger}\overleftrightarrow{D_{\mu}}w,
\end{align}
where the last term is the $w$-boson current
$i[w^{\dagger}(\partial_{\mu}-iA_{\mu})w-w(\partial_{\mu}+iA_{\mu})w^{\dagger}]$.
These equations of motion indicate that one flux quanta of
$a_{\mu}$($A_{\mu}$) carries two units of $A_{\mu}$($a_{\mu}$) gauge
charge. And one unit of $A_{\mu}$($a_{\mu}$) gauge charge in turn
carries $\pi$-flux of $a_{\mu}$($A_{\mu}$) gauge field. As a result,
$f$-spinon sees the $w$-boson as a $\pi$-flux and vice versa. These
are the two gapped fundamental excitations in a $Z_2$ gauge theory:
gauge charge and vison, and they are dual degrees of freedom. If we define the $f$-spinon ($w$-boson) as the $Z_2$ gauge charge,
$w$-boson($f$-spinon) will be the vison. From the viewpoint of $f$-spinons, we
just showed that their visons, \ie $w$-bosons carry spin
quantum number.

Now let's discuss the monopoles of $a_{\mu}$ and $A_{\mu}$ in
Eq.(\ref{eq:eff_lag}). In $\hat n$ ordered phase (CAF phase), where
$A_{\mu}$ is Higgsed out, the $a_{\mu}$ monopole is bound with a
spin flip due to the QSH effect. Therefore $a_{\mu}$ monopole events
are suppressed in the CAF phase by spin rotation symmetry and
$a_{\mu}$ stays in Coulomb phase. However in the $\hat n$ disordered
phase, there is no QSH effect and monopole events of both $a_{\mu}$
and $A_{\mu}$ are allowed. Let us denote their monopole creation
operators as $V_{a}^{\dagger}$ and $V_{A}^{\dagger}$ respectively.
What are the physical consequences of these monopole events?

By equations of motion it is shown that an $a_{\mu}$ monopole
creates 2 units of $A_{\mu}$ gauge charge, and vice versa. Thus such
monopole events simply mean that
$f_{\alpha}^{\dagger}f_{\beta}^{\dagger}$ and
$w_{\alpha}^{\dagger}w_{\beta}^{\dagger}$ pairing terms exist in the
effective action. These pairing terms, or monopole events, break the
$U(1)$ gauge symmetry down to $Z_2$.

We should focus on one set of dual variables $f$($V^{\dagger}_{A}$)
and $w$($V^{\dagger}_{a}$) to write down the effective theory. First
take a look at the $V_{A}^{\dagger}$ operator: what is the symmetry
of the corresponding $f$-spinon pairing term? The quantum number of
this monopole operator is determined by the WZW term: a Skyrmion of
$\hat n$ carries an s-wave pair of $f$-spinon. Therefore we showed
that $\hat n$ disordered phase is nothing but the SPS, whose
mean-field ansatz is given in Eq.(\ref{eq:SPS_mf}).

The phase transition from SPS to CAF phase is described by $w$-boson
condensation in Eq.(\ref{eq:eff_lag}). But due to the topological
Berry's phase, this is a rather "high-energy" description of the
criticality. To find a low energy description without topological
terms, we should resort to another formulation as shown in the next
section.

\textbf{Duality between Schwinger-fermion and Schwinger-boson
representations}

In this section we focus on the dual variables of $f$-spinons: the
$w$-bosons. The SPS phase is then a $Z_2$ phase with $w$-bosons as
$Z_2$ charges, but $f$-spinons as visons. In this formulation
SPS-CAF phase transition is naturally presented as a Higgs
condensation of $w$-bosons.

First we need to represent the order parameters of the CAF phase in
terms of $w$. 
The QSH order is $\hat n= w^{\dagger}\vec \sigma w$, but what is the
Neel order parameter? Neel order in CAF phase corresponds to the
monopole of $a_{\mu}$, namely a pairing of $w$-boson. There are two
spin-1 bosonic pairing order parameters satisfying this requirement,
\ie the real and imaginary part of
$(i\sigma_yw^*)^{\dagger}\vec\sigma w$:
\begin{align}
 \hat n_1+i\hat n_2=(i\sigma_yw^*)_{\alpha}^{\dagger}\vec\sigma_{\alpha\beta} w_{\beta}.
\end{align}
It is easy to verify that $\hat n=\hat n_1\times \hat n_2$, so there
are only two independent vectorial order parameters. The issue is,
which one is the Neel order parameter $\vec N$: $\hat n_1$ or $\hat
n_2$?

A $U(1)$ gauge transformation $w\rightarrow e^{i\theta}w$ generates
a rotation in the $\hat n_1,\hat n_2$ plane. By fixing a proper
gauge, we can always choose $\hat n_1$ as the Neel order. We will
work within this gauge $\vec N=\hat n_1$ throughout the phase
transition. Such a gauge fixing breaks the $U(1)$ gauge redundancy
down to $Z_2$: $w\rightarrow \pm w$.

The physical symmetries of the QSH (or vector spin chirality) and
the Neel order parameters completely determine the transformation
rules of the $w$-boson up to a $Z_2$ gauge redundancy:
\begin{align}
T_1,T_2: &&&\hat n\rightarrow \hat n,&& \vec N\rightarrow \vec N,&&w\rightarrow w,\notag\\
\bs{T}: &&&\hat n\rightarrow \hat n,&& \vec N\rightarrow -\vec N,&& w\rightarrow iw^{*},\notag\\
\sigma: &&&\hat n\rightarrow -\hat n,&&\vec N\rightarrow -\vec N,&& w\rightarrow i\sigma_y w^*,\notag\\
C_6: &&&\hat n\rightarrow \hat n,&&\vec N\rightarrow -\vec
N&&w\rightarrow iw.\label{eq:sym}
\end{align}
where time-reversal transformation $\bs{T}$ is anti-unitary. The
reason why there are no further arbitrariness on the transformation
rules of $w$ can be easily understood by the following construction.
If we write $w$-boson as an SU(2) matrix:
\begin{align}
 U=\begin{pmatrix}w_1&w_2^*\\w_2&-w_1^*\end{pmatrix},
\end{align}
then the most general $O(4)$ transformation leaving
$|w_1|^2+|w_2|^2=1$ is $U\rightarrow V_LUV_R$, where $V_L$ and $V_R$
are both $SU(2)$ rotations ($V_L$ is spin rotation), and $O(4)\sim
SU(2)_L\times SU(2)_R$. In this representation, the vectors $ \hat
n_1,\hat n_2,\hat n$ are the 1st, 2nd and 3rd columns of a 3 by 3
rotation matrix $R$:\cite{Lee1998}
\begin{align}
 R^{ab}=\frac{1}{2}\mbox{Tr}(U^{\dagger}\sigma^aU\sigma^b)\label{eq:Rab}
\end{align}
Clearly, to leave $R$ invariant, the transformations $V_{L,R}$ must
be $\pm 1$.

These symmetry transformation rules allow us to reveal the
connection between the SPS state here and the 0-flux state in the
Schwinger-boson representation obtained by Wang\cite{Wang2010}. In
Wang's work, the Neel order is represented by the $z$-boson as $\vec
N= z^{\dagger}\vec \sigma z$ in the effective theory. From
Eq.(\ref{eq:Rab}), we can easily construct the duality
transformation between the $w$-boson and $z$-boson representations:
$U_w=U_zV_R$, $V_R=e^{i\frac{\pi}{4}\sigma_y}$, namely:
\begin{align}
 w&=\frac{1}{\sqrt{2}}(z-i\sigma^y z^*)&\mbox{or}&& z&=\frac{1}{\sqrt{2}}(w+i\sigma^yw^*)\label{eq:duality}
\end{align}
Under duality transformation:
\begin{align}
 \vec N&=\mbox{Re}[(i\sigma_yw^{*})^{\dagger}\vec \sigma w]=z^{\dagger}\vec \sigma z,\notag\\
\hat n&=w^{\dagger}\vec \sigma
w=-\mbox{Re}[(i\sigma_yz^{*})^{\dagger}\vec \sigma z].
\end{align}
From Eq.(\ref{eq:sym}),(\ref{eq:duality}), we can obtain
transformation rules of $z$-bosons:
\begin{align}
T_1,T_2: &&& z\rightarrow z,\notag\\
\bs{T}: &&&  z\rightarrow \sigma_y z,\notag\\
\sigma: &&&  z\rightarrow i\sigma_y z^*,\notag\\
C_6: &&& z\rightarrow \sigma_y z^*,\label{eq:zsym}
\end{align}
which are exactly the transformation rules found by
Wang\cite{Wang2010} for 0-flux state up to a $Z_2$ gauge
arbitrariness. This explicitly confirms that the $z$-bosons
constructed in Eq.(\ref{eq:duality}) are the same $z$-bosons
discussed by Wang, and the SPS phase here is identical to the 0-flux
phase in Schwinger-boson description.

Following the discussion in \Ref{Wang2010}, we can write down the
general symmetry-allowed effective theory for the phase transition
in terms of $z$-boson:
\begin{align}
 L=&|\partial_{\tau}z|^2+c^2|\nabla z|^2+m^2|z|^2+u (|z|^2)^2\notag\\
&+\lambda_H \;(i\sigma_y z^*)^{\dagger}\big[\sum_i(\vec d_i\cdot
\nabla)^3\big] z+h.c.\label{eq:eff_z_boson}
\end{align}
For instance, the single time derivative term
$z^{\dagger}\partial_{\tau} z$ is forbidden by $\sigma$, and
$z^T(-i\sigma_y)\partial_{\tau} z$ is forbidden by $C_6$. Here
$\lambda_H$ is the Higgs coupling which reduces the $U(1)$ gauge
degrees of freedom in the $z$-boson formulation down to $Z_2$.
$\vec d_1=-\vec a_1$,$\vec d_2=\vec a_2$, and $\vec d_3=\vec
a_1-\vec a_2$ as shown in FIG. \ref{fig:honeycomb}.  The Higgs coupling can also be written as a
pairing of $w$-bosons: $\lambda_H \;(i\sigma_y
w^*)^{\dagger}\big[\sum_i(\vec d_i\cdot \nabla)^3\big] w+h.c$. By
naive power counting $\lambda_H$ is irrelevant, therefore we have an
$O(4)$ critical point between the CAF ($z$-condensed) phase and the
SPS ($z$-gapped) phase. The critical behavior of this transition is
well-studied\cite{Chubukov1994,chubukov1994a,Isakov2005}.

\textbf{Discussion} In this study, our main prediction is the CAF
phase. Unlike the usual AF phase, CAF phase has two order
parameters: Neel $\vec N$ and QSH $\hat n$. As CAF phase is very
likely to be the magnetic ordered phase adjacent to the spin liquid
phase found in the Hubbard model on the honeycomb lattice, in the
following we propose an explicit numerical methods to detect the CAF
phase.

One can directly measure the QSH order by $\langle \vec
n(x)\cdot\vec n(0)\rangle$ correlation function, or the vectorial
spin chirality correlation function $\langle(\nu_{i+x,j+x}\vec
S_{i+x}\times\vec S_{j+x})\cdot(\nu_{ij}\vec S_i\times\vec
S_j)\rangle$. Because QSH order is odd under $\sigma\cdot\bs T$,
while pure Neel order is $\sigma\cdot\bs T$ even, one does not
expect a long range correlation of QSH order in a usual AF phase.
Thus the long range QSH correlation function is an intrinsic
signature of the CAF phase. In addition, one can show that the QSH
direction is normal to the Neel direction. For example, one can pin
the Neel order by an infinitesimal (in thermodynamic limit)
staggered magnetic field along $S_z$ direction, and then to measure
the QSH order parameters by correlation function. One should find
the QSH order parameters have only $x,y$ components. In the real
world, as mentioned in \Ref{Meng2010}, such an exotic spin liquid
may be realized in many candidate systems: \eg expanded
graphene-like system in group $\Rmnum4$
elements\cite{Seehofer1993,Cahangirov2009}, as well as fermions in
optical lattices\cite{Duan2003,Jordens2008}.

YR thanks for helpful comments from Ashvin Vishwanath. YML thanks
Prof. Ziqiang Wang for support during this work under DOE Grant
DE-FG02-99ER45747. YR is supported by the startup fund at Boston
College.


\appendix
\section{Derivation of the mutual Chern-Simons term}\label{app:mutual_CS}

We start from the following low energy effective Lagrangian of
spinon fields $\psi$ (see Eq. \ref{eq:CAF_eff}) in imaginary time
(\ie Euclidean space-time):
\begin{eqnarray}
\mathcal{L}_{eff}=\bar\psi\big[\imth\gamma_\mu(\partial_\mu-\imth
a_\mu)+\imth m~\hat n\cdot\vec\sigma~\big]\psi
\end{eqnarray}
where we define $\bar\psi\equiv\psi^\dagger\gamma^0$. For simplicity
let's denote $-\imth\mathcal{G}^{-1}=\gamma_\mu(\partial_\mu-\imth
a_\mu)+m\hat n\cdot\vec\sigma$, then integrating out spinon fields
$\psi$ yield the effective action
$\mathcal{S}=-\ln\det(\mathcal{G}^{-1})=-\text{Tr}\ln(\mathcal{G}^{-1})$.
Following the spirit of Abanov and
Wiegmann\cite{Abanov2000,Abanov2001}, we use large-$m$ expansion to
obtain the low energy effective theory in the longwavelength limit
$\omega\ll m$. By defining
$\mathcal{G}_0^{-1}=\imth(\gamma_\mu\partial_\mu+m\hat{n}\cdot\vec\sigma)$
we have $\mathcal{G}^{-1}=\mathcal{G}_0^{-1}+a_\mu\gamma_\mu$. Let's
denote $\plash{\partial}\equiv\gamma_\mu\partial_\mu$ and similarly
$\plash{a}\equiv\gamma_\mu a_\mu$ and we have
\begin{eqnarray}
&\notag\mathcal{S}=-\text{Tr}\ln(\mathcal{G}_0^{-1}+\plash{a})=-\text{Tr}\ln(\mathcal{G}_0^{-1})-\text{Tr}\ln(1+\mathcal{G}_0\plash{a})\\
&=\mathcal{S}_0+\sum_{l=1}^\infty(-1)^l\text{Tr}(\mathcal{G}_0\plash{a})^l\notag
\end{eqnarray}
Here $\mathcal{S}_0$ gives the nonlinear-sigma-model dynamics
$\sim(\partial_\mu\hat{n})^2$ of vector $\hat{n}$, while the
coupling between vector $\hat{n}$ and gauge field $a_\mu$ is given
by the 2nd term. In the large-$m$ expansion we consider only the
leading-order term:
\begin{eqnarray}
\mathcal{S}_1=-\text{Tr}(\mathcal{G}_0\plash{a})=-\text{Tr}\Big\{\mathcal{G}_0^{-1}\big[\mathcal{G}_0^{-1}(\mathcal{G}_0^{-1})^\dagger\big]^{-1}\plash{a}\Big\}\notag
\end{eqnarray}
It's straightforward to check that
$\mathcal{G}_0^{-1}(\mathcal{G}_0^{-1})^\dagger=-\partial^2+m^2-m\vec\sigma\cdot\plash\partial\hat
n$, therefore large-$m$ expansion leads to
\begin{eqnarray}
\big[\mathcal{G}_0^{-1}(\mathcal{G}_0^{-1})^\dagger\big]^{-1}=(-\partial^2+m^2)^{-1}\sum_{l=0}^\infty\big(\frac{m\vec\sigma\cdot\plash{\partial}{\hat
n}}{-\partial^2+m^2}\big)^l\notag
\end{eqnarray}
and consequently
\begin{eqnarray}
\mathcal{S}_1=-\sum_{l=0}^\infty\text{Tr}\Big\{\frac{\imth(\plash\partial+m\hat{n}\cdot\vec\sigma)}{-\partial^2+m^2}\big(\frac{m\vec\sigma\cdot\plash{\partial}{\hat
n}}{-\partial^2+m^2}\big)^l\plash{a}\Big\}\notag
\end{eqnarray}
It turns out that $l=0,1$ terms both vanish and the leading-order
correction to the low energy effective action is the following
topological term:
\begin{eqnarray}
&\nonumber\mathcal{S}_{topo}=-a_\mu\text{Tr}\Big[\gamma_\mu\frac{\imth
m\hat{n}\cdot\vec\sigma}{-\partial^2+m^2}(\frac{
m\gamma_\nu(\partial_\nu\hat{n})\cdot\vec{\sigma}}{-\partial^2+m^2})^2\Big]\\
&=\frac\imth{4\pi}\epsilon_{\mu\nu\lambda}~a_{\mu}~\hat{n}\cdot(\partial_\nu\hat{n}\times\partial_\lambda\hat{n})\label{eq:topo_term}
\end{eqnarray}

Notice that in the $CP^1$ parametrization of order parameter
$\hat{n}=w^\dagger\sigma w$,  spinor $w$ is the eigenvector of
$\hat{n}\cdot\vec\sigma$ whose spin orientation is along unit vector
$\hat{n}$. Therefore the Skyrmion current of $\hat{n}$
\begin{eqnarray}
J_\mu^{Sk}=\frac12\cdot\frac1{4\pi}\epsilon_{\mu\nu\lambda}\hat{n}\cdot(\partial_\nu\hat{n}\times\partial_\lambda\hat{n})
\end{eqnarray}
which equals half the winding number of $\hat{n}$ wrapping around
$S^2$, is nothing but the Berry's phase\cite{Berry1984} for spinor
$w$. Since spinor $w$ obtains $\pi$ phase (\ie a minus sign) as
$\hat{n}$ wraps around $S^2$ once (\ie $\hat{n}$ covers $4\pi$ solid
angle), this gives a direct correspondence between the Skyrmion
current density and the $U(1)$ gauge field strength $F_{\mu\nu}$
coupled to $w$:
\begin{eqnarray}
\frac1{4\pi}\epsilon_{\mu\nu\lambda}\hat{n}\cdot(\partial_\nu\hat{n}\times\partial_\lambda\hat{n})=\frac1{\pi}\epsilon_{\mu\nu\lambda}\partial_\nu
A_{\lambda}
\end{eqnarray}
Therefore the topological term (\ref{eq:topo_term}) is exactly the
mutual Chern-Simons term mentioned in (\ref{eq:eff_lag})
\begin{eqnarray}
\nonumber\mathcal{S}_{topo}=\frac{\imth}{\pi}\epsilon_{\mu\nu\lambda}~a_{\mu}~\partial_\nu
A_{\lambda}
\end{eqnarray}

\end{document}